\documentclass[12pt,preprint]{aastex}

\newcommand{\arcsecpt}{\makebox[0pt]{ .}^{\prime\prime}}
\newcommand{\arcminpt}{\mbox{\,\makebox[0pt]{.}^{\prime}}}

\newcommand{\HI}{\mbox{\normalsize H\thinspace\footnotesize I}}

\newcommand{\lya}{\mbox{\,Ly$\alpha$}}

\newcommand{\kmsec}{\mbox{\,km s$^{-1}$}}
\newcommand{\NV}{\ion{N}{5}}
\newcommand{\CIV}{\ion{C}{4}}
\newcommand{\MgII}{\ion{Mg}{2}}

\newcommand{\OVI}{\ion{O}{6}}
\newcommand{\OVII}{\ion{O}{7}}
\newcommand{\OVIII}{\ion{O}{8}}

\begin{document}

\title{Discovery of Associated Absorption Lines in
an X-Ray Warm Absorber: 
{\it HST-FOS} Observations of MR2251$-$178\footnote{Based on Observations with the NASA/ESA {\it Hubble Space Telescope}, obtained at the Space Telescope Science Institute, which is operated by AURA, Inc., under NASA contract NAS 5-26555.}}

\author{Eric M. Monier\footnote{monier@astronomy.ohio-state.edu}\.\ and Smita Mathur\footnote{smita@astronomy.ohio-state.edu}}
\affil{Department of Astronomy, The Ohio State University}
\affil{ 140 W. 18th Ave., Columbus, OH  43204}
\and
\author{Belinda Wilkes and Martin Elvis}
\affil{Harvard-Smithsonian Center for Astrophysics}
\affil{60 Garden Street, Cambridge, MA 02138}

\begin{abstract}

The presence of a ``warm absorber'' was first suggested to explain 
spectral variability in an X-ray spectrum of the radio-quiet QSO MR2251$-$178.
A unified picture, in which
X-ray warm absorbers and ``intrinsic'' UV absorbers are the same,
offers the opportunity to probe the nuclear environment
of active galactic nuclei.  To test this scenario and understand the
physical properties of the absorber, we obtained a UV spectrum of
MR2251$-$178 with the Faint Object Spectrograph (FOS) onboard
the {\it Hubble Space Telescope (HST)}.  The {\emph HST}
spectrum clearly shows absorption due to Ly$\alpha$, \ion{N}{5}
and \ion{C}{4}, blueshifted by 300 \kmsec\ from the emission redshift
of the QSO.  The rarity of both X-ray and UV absorbers in radio-quiet
QSOs suggests these absorbers are physically related, if not identical.
Assuming the unified scenario, we place constraints on the
physical parameters of the absorber and conclude the mass outflow
rate is essentially the same as the accretion rate in MR2251$-$178.

\end{abstract}

\keywords{quasars: absorption lines ---
quasars: individual (MR2251$-$178) -- ultraviolet: galaxies}

\newpage

\section{Introduction}

Absorption due to \OVII\ and \OVIII\ in the X-ray and \OVI\
in the UV can in some cases be  modeled  
by a single UV/X-ray absorber (Mathur 1994; Mathur, Wilkes, \& Elvis 
1998), offering a unique way to determine the physical conditions 
of the absorbing material in active galactic nuclei (AGNs).
As more X-ray/UV absorbers have been found (e.g., Mathur 1994; Mathur, Elvis, 
\& Wilkes 1995; Mathur, Elvis, \& Singh 1995; Mathur, Wilkes, \& Aldcroft
1997; Shields \& Hamann 1997; Mathur et al. 1998; Hamann, Netzer,
\& Shields 2000) it has become clear these share common characteristics.
Though not all of these absorbers can be readily modeled by a single zone, 
all are composed of highly ionized, low density, high
column density, outflowing gas sitting outside the broad emission line region
(BELR).  The X-ray/UV absorbers are therefore an important nuclear component
of AGNs, representing a wind or outflow carrying away significant kinetic
energy at a mass-loss rate comparable to the accretion rate needed
to power the AGN (Mathur et al. 1995).  The properties of these absorbers
can be invoked in AGN models developed to explain such outflows 
(see Elvis 2000 and references therein), and will aid in expanding 
such models into a wider unification scenario (Elvis 2000).

It remains to be determined how common X-ray/UV absorbers are.  In a sample of
eight Seyfert galaxies, Crenshaw et al. (1999) found six of them showed
both UV and X-ray absorption, strengthening the UV/X-ray connection. 
The question remains whether X-ray absorbers are physically 
related to UV absorbers in all AGN (two of the Seyferts of the Crenshaw et al.
sample were not X-ray warm absorbers and showed no UV absorption), or, if not,
in what fraction of cases does the unified model apply. 
 To help answer these questions, we initiated an {\it HST} program
to search for UV absorption in two radio-quiet QSOs with 
known X-ray warm absorbers.  

Mathur et al. (1998) reported UV absorption in the first of these
objects, PG1114+445.  Here we present the detection of UV absorption
in our second candidate, MR2251$-$178, a low-redshift ($z_{em} = 0.0640$)
radio-quiet QSO.  
In addition to being the first X-ray selected QSO (Ricker et al. 1978), 
MR2251$-$178 was the object for which Halpern (1984) originally suggested
the presence of partially ionized, optically thin material: the X-ray
warm absorber.

In {\S 2} we describe the HST observations and the analysis of the UV absorption
lines.  Section 3 relates the UV absorption to the X-ray absorption seen in
ASCA data analyzed by Reynolds (1997) and presents some simple physical quantities
based on the assumption that the UV and X-ray absorbers are the same.  We present
our conclusions in {\S 4}.

\section{The {\it HST} spectra}

On 1996 February 2, MR2251$-$178 was observed using three  gratings of
the {\it Faint Object Spectrograph} onboard {\it HST}.  The target
was centered in the 1'' circular aperture through a four step peak-up
sequence to ensure a pointing accuray of 0$\arcsecpt$12.  Two exposures
totalling 5740s were obtained using the G130H grating and blue side of the
detector.  The exposures using the red side and the G190H and G270H gratings
were 1730s and 300s, respectively.

The data were reduced using the standard pipeline processing and calibration 
files, and the individual G130H spectra were combined.  The final spectra
are shown in Figure 1.  Absorption lines of \lya, \CIV\ and \NV\ can be clearly
seen, as predicted from the X-ray/UV models.  The strong absorption lines of
\lya\ and \CIV\ are superposed on the emission lines of \lya\ and \CIV.  Weaker
absorption due to the \NV\ doublet is seen against the continuum; no
measurable \NV\ emission is present.
The spectra do not show absorption due to low-ionization \MgII\, 
indicating the X-ray absorber is highly ionized. 

There is an inherent uncertainty involved in measuring an absorption line 
lying on top of a broad emission line profile because the unabsorbed shape
of the emission line is unknown.  We therefore used two methods to 
measure the line equivalent widths (EWs) as in Mathur et al. (1998).  
We made an initial measurement using the 
IRAF SPLOT task to fit Gaussians across the absorption in the emission line
profiles.  A single Gaussian was used for \lya\ and \NV\ lines, and 
the \CIV\ doublet is deblended into two Gaussians.  Results for the EWs 
and FWHMs are listed in Table 1.  We estimate the errors on the EWs for
\lya\ and \CIV\ to be $\sim$40\%.  The continuum is more readily determined
on either side of the \NV\ lines, hence the error on the SPLOT measurement
is estimated at $\sim$10\%.  The \CIV\ and \NV\ doublet ratios are both 
$\approx$1.7, so the doublets may be partially saturated, though
not as severely as in PG1114+445.  Higher resolution spectra would be needed
to address the degree of saturation.  Also, though it is probable the individual
lines will break into multiple components at higher resolution
(Ly$\alpha$ in fact has two components; see \S 3.1), we treat them as single lines here.

To obtain a more accurate estimate of the absorption line EWs, we used the SPECFIT
(Kriss 1994) task under STSDAS to simultaneously fit the underlying continuum
and the emission and absorption line profiles.  The continuum was
characterized by a simple power law, the emission lines were fit with
multiple Gaussians, and each absorption line was fit with a single Gaussian.
First, the power-law continuum was fit to featureless parts of the spectrum
on either side of the emission lines and the absorption to be measured.
The parameters of the power-law continuum -- the slope and the normalization 
-- were then fixed in subsequent fits.  The parameters of the Gaussians
used to fit the emission lines (flux, centroid, FWHM and skew) and the
absorption lines (flux, centroid and FWHM) were allowed to vary freely.
The fits to the \lya, \NV\ and \CIV\ emission/absorption profiles are
shown in Figure 2, and the corresponding EW and FWHM of the absorption
fits are given in Table 1.  The results of the SPLOT and SPECFIT methods
are qualitatively similar, although the SPECFIT values are systematically
larger due to the higher continuum provided by fitting the emission
lines.  These differences are unimportant to the discussion.

The observed FWHM of the absorption lines is 1.3 - 3.0 \AA\ 
($>$300 km s$^{-1}$) with
SPLOT and 1.4 - 3.3 \AA\ ($>$320 km s$^{-1}$) with SPECFIT.  For each ion the total 
absorption is likely resolved given the $\approx$230 \kmsec\ resolution of FOS, so
the absorber is dispersed in velocity space.  In a higher resolution spectrum,
however, the absorption lines may split into multiple components.
  The average absorption redshift
for the five lines is $<$z$>$=0.063, blueshifted by 300 \kmsec\ from the
QSO emission redshift of $z_{em} = 0.0640$ (Bergeron et al. 1983).
We expect that \OVI\ absorption is also present, but it lies below
the blue cutoff of the G130H grating at the redshift of MR2251$-$178.

MR2251$-$178 was also observed in 1998 Dec
by J. Stocke with HST-STIS over the wavelength range 1250--1300\AA.  
The \lya\ absorption line breaks into a broad component and a narrow component
at the $\approx$12 \kmsec\ resolution of the STIS G140M grating.
Figure 3 shows the \lya\ region and a fit to the absorption profile 
produced with SPECFIT as described previously.
The resulting line centers, EWs and FWHMs of the two components are
listed in Table 1.

\section{The X-Ray/UV Absorber}

High-ionization UV absorption lines due to \lya, \NV, and 
\CIV\ are seen in the {\it HST} spectrum of MR2251$-$178, as predicted 
by UV/X-ray models.  Mathur et al. (1998) estimate the chance probability that
a radio-quiet QSO will have both associated
UV and X-ray ionized absorption to be $\sim 1.7 \times 10^{-3}$.
This strongly suggests the two absorption systems are physically
related.

\subsection{Physical Properties of the Ionized Gas}

To explore the possibility that the UV and X-ray absorption lines arise
in the same component of the nuclear material of the AGN, we need to compare
the properties of the UV and X-ray absorbers.

We took the measurements of the X-ray absorber in MR2251$-$178 
from the analysis of ASCA data (obtained 6 Nov 1993) by Reynolds (1997; Komossa 2001 obtains
a similar result using ROSAT data from the same epoch).  The total equivalent
hydrogen column density of the X-ray absorber is $N_{\rm H} = 
 5 \times 10^{21}$ atoms cm$^{-2}$, determined 
by fitting the warm absorber with a one-zone model using the 
CLOUDY photoionization code (Ferland 1996).  Reynolds constrained the 
optical depths of the \OVII/\OVIII\ edges in a simple two-edge model with
a power-law continuum ($\alpha = 1.73$).
We used these values as an input to CLOUDY 
to derive the column densities of the UV ions seen in absorption in the {\it HST} data.
The standard ``Table AGN'' (Mathews \& Ferland 1987) continuum 
(with a resulting ionization parameter of $U = 1.0.)$\footnote{{\it U} is defined as the dimensionless ratio of
ionizing photon to hydrogen number density.}
was used with the assumption of solar abundances.  The shape of the observed
UV continuum of MR2251$-$178 (Figure 1) is consistent with that in "Table AGN."
Table 1 lists the CLOUDY predictions for the ionic column densities of \ion{H}{1},
\ion{N}{5}, and \ion{C}{4}, as derived from the X-ray data. 

To compare these predictions to what is observed in the UV, it is necessary
to convert the equivalent widths to column densities.  As discussed
in $\S$2, the lines may be somewhat saturated, thus the column
densities can be calculated only if the velocity dispersion parameter
($b$ parameter) of the lines is known. 
If the lines are resolved, then $b{\approx}$ 450 \kmsec\ [$b = $FWHM/(4ln2)$^{1/2}$;
the FWHM is given in Table 1] for 
\lya\ and $\approx$ 200 - 270 \kmsec\  for \CIV\ and \NV.  
The X-ray prediction for the column density of hydrogen, $N_{\rm HI} = 8 \times 10^{14}$ cm$^{-2}$,
is therefore about a factor of 2 larger than the minimum value measured in the FOS data
($N_{\rm HI} \approx 4 \times 10^{14}$ cm$^{-2}$).
Agreement between the UV and X-ray values is achieved 
for $b \approx 200$ \kmsec, implying the line breaks into at least two components at 
higher resolution, as is shown in the STIS data (\S 2).\footnote{A weak feature at $\approx$ 1295.5 \AA\
may also be \lya\ redshifted relative to the quasar, as has been 
observed in \ion{C}{4} lines of NGC 5548 (Mathur, Elvis, \& Wilkes 1999);
this feature would not contribute significantly to $N_{\HI}$.}
 A curve-of-growth analysis using the new $b$-values
($\sim$65 and $\sim$280 \kmsec) results in a combined
total \lya\ column density of $\sim7 \times 10^{14}$, in
good agreement with the X-ray prediction for \ion{H}{1}.
 
The X-ray predicted values for the \ion{N}{5} and \ion{C}{4} are
smaller than those calculated from the UV equivalent width measurements.
For \ion{N}{5}, the X-ray prediction is $N_{\rm NV} = 1.6 \times 10^{14}$ cm$^{-2}$,
while the UV equivalent width measurement gives $N_{\rm HI} = 8 \times 10^{14}$ cm$^{-2}$,
assuming the lines are resolved.  Similarly for \ion{C}{4}, the X-ray prediction 
of $N_{\rm CIV} = 1.6 \times 10^{13}$ cm$^{-2}$ is more than an order of magnitude
smaller than the UV measurement of $N_{\rm CIV} \ge 7 \times 10^{14}$ cm$^{-2}$.
The X-ray predictions for $N_{\rm NV}$ and especially $N_{\rm CIV}$
lie on the linear part of the curve of growth. 
If the UV and X-ray absorbers are the same, we predict these absorbers should
break into multiple components -- each with lower $b$ -- in UV data of higher resolution, 
leading to values in closer agreement with the X-ray predictions.
Kinematically complex UV absorption lines have already been seen in high-resolution
data of Mrk 509 (Crenshaw, Boggess \& Wu 1995), NGC 4151
(Weymann et al. 1997), NGC 3516 (Kriss et al. 1996, Crenshaw et al. 1999),
and NGC 5548 (Mathur et al. 1999; Crenshaw \& Kraemer 1999).

Also, note the column densities derived here from the
X-ray data are based on the Table AGN continuum.  The shape of the actual MR2251$-$178
continuum in the unobserved EUV region may greatly affect the predicted
column densities (Mathur et al. 1995).  A better match between the observed
\& predicted values may be obtained if the spectral energy distribution of
MR2251$-$178 has a different shape.
In any case, the X-ray absorber is a large contributor to the
absorption measured for \CIV\ and \NV\ in the UV, providing support
for the idea that the UV and X-ray absorption is occurring in the same material.

\subsection{Variability}

Halpern (1984) found that the warm absorber in MR2251$-$178 was highly variable.
ASCA data of MR2251$-$178 obtained in 1996 -- near the epoch of the HST-FOS data -- show 
a decrease in X-ray intensity from log $\xi$ = 1.35 in 1993 to log $\xi$ = 0.96 in 1996
(Otani, Kii, \& Miya 1998), where $\xi$ is an ionization parameter defined as $\xi \equiv L/nR^2$. 
If the ionization parameter $U$ varied similarly, 
then in 1996 $U = 0.61$ and the X-ray-predicted column densities would be
$N_{\rm CIV} = 3.9 \times 10^{14}$ cm$^{-2}$ and 
$N_{\rm NV}  = 2.1 \times 10^{15}$ cm$^{-2}$, closer to the values measured in
the UV data.  In this case, $N_{\rm HI}$ grows to $3.4 \times 10^{15}$ cm$^{-2}$.

\subsection{Physical Implications}

As noted, the UV absorption indicates the material is flowing
outward from the UV/X-ray absorber at $\approx$300 \kmsec.  With the assumption
that the UV and X-ray absorbers are the same we can determine the physical conditions
and mass outflow rate of the absorbing material.
We use the total hydrogen
column of $N_{\rm H}$ = 5$\times 10^{21}$ cm$^{-2}$ and the ionization 
parameter $U = 1.0$ from the CLOUDY modeling of the X-ray data.  The distance of the absorber from
the nucleus is constrained by $U = (Q/4{\pi}r^2n_Hc)^{1/2}$, where $Q$ is
the rate of ionizing photons.  We used the HST data at 1339 \AA\ rest wavelength 
to scale the standard Table AGN continuum to the luminosity of MR2251$-$178
($L_{bol} = 5 \times 10^{45}$ ergs s$^{-1}$).
Integrating $L_{\nu}/(h{\nu})$ over ${\nu} \ge 13.6$eV 
gives $Q = 6h^{-2} \times 10^{54}$ s$^{-1}$, where $h$ is the Hubble
constant in units of 100 \kmsec\ Mpc$^{-1}$.  This puts the distance of
the absorber from the nucleus at $r = 3.0 \times 10^{18}n_5^{-1/2}$ cm (where $H_{\rm o} = 75$
\kmsec\ Mpc$^{-1}$ and $n_5$ is the density in units of $10^5$ cm$^{-3}$).

The absorption extends below the continuum level (Figure 1), and so must cover a
substantial fraction of the BELR as well as the continuum source.  
The absorber may lie outside the BELR or it
could also be cospatial with it if most of the emission is beamed
from the far side of the BELR.  The size of the BELR in MR2251$-$178 
can be estimated by scaling the BELR of NGC 5548 ($L_{bol} = 5 \times
10^{44}$ ergs s$^{-1}$) by $L^{1/2}$ (Davidson \& Netzer 1979; but see
Kaspi et al. 2000).  The distance -- determined from reverberation mapping --
of the BELR from the central continuum in NGC 5548 is $r \simeq 10$
light-days (Clavel et al. 1991), so the absorber in MR2251$-$178 must be at least $\sim$32
light-days, or $r \ge 8.2 \times 10^{16}$ cm from the central source.
The corresponding upper limit on the density is then $n < 1.34 \times 10^8$   
cm$^{-3}$.  To check the radius estimate, we specified as an input to CLOUDY the luminosity
of MR2251$-$178 at a rest wavelength of 1339 \AA, log(${\nu}L_{\nu}$) = 43.78,
and varied the input radius to obtain the ionization parameter output
previously.  The radius at which $U = 1.0$ was $r = 8.76 \times 10^{16}$ cm.
For a covering factor of $f = 0.1$, the column density and particle density
of the aborber imply a mass of $M = 47f_{0.1}n_5^{-1} M_{\odot}$.
The mass outflow rate (Mathur et al. 1995) would then be 
$\dot{M}_{out} = 0.9f_{0.1} M_{\odot}$ yr$^{-1}$,
the same rate needed to power MR2251$-$178 at 10\% efficiency.
The rate of kinetic energy carried away in the flow is $\dot{M}v_{out}^2/2 = 
8 \times 10^{40}$ ergs cm$^{-2}$ s$^{-1}$.

\section{Conclusions}

The {\it HST} FOS UV spectrum of MR2251$-$178 exhibits associated
high-ionization, UV absorption lines, as predicted from models of the
X-ray ionized absorber.  The X-ray/UV absorber is situated outside
the BELR or cospatial with it and outflowing with a line-of-sight
velocity of $\approx$300 \kmsec.  The mass outflow rate of 0.9 M$_{\odot}$
yr$^{-1}$ for a 10\% covering factor is equivalent to the accretion rate
onto the nuclear black hole.

We expect future higher resolution observations 
will reveal each of the \CIV\ and \NV\ absorption lines 
breaking into at least two components.

The rarity of both UV and X-ray absorbers individually in radio-quiet QSOs
virtually requires that the X-ray and UV absorbers are closely 
physically related.  The consistency of the column densities obtained
from both UV and X-ray data suggests the absorbers are perhaps identical.
At a minimum, the X-ray absorber makes a substantial contribution to the
absorption seen in the UV.  Thus, the absorber in MR2251$-$178 satisfies
both statistical as well as physical tests of our X-ray/UV absorber
model.

\figcaption[hst.ps]{{\it HST} spectrum of MR2251$-$178 showing absorption
lines of Ly$\alpha$, \ion{N}{5}, and \ion{C}{4}.  The lower line on each
panel is the error spectrum.}

\figcaption[fits.ps]{SPECFIT fits to the absorption lines of Ly$\alpha$ and \ion{N}{5}
(upper panel) and \ion{C}{4} (lower panel).  Results of the fits are shown in 
Table 1.}

\figcaption[stis.ps]{SPECFIT fit to the {\it HST-STIS} spectrum of the
\lya\ absorption line of MR2251$-$178.  Results of the fit are shown in Table 1.}

\begin{figure}
\plotone{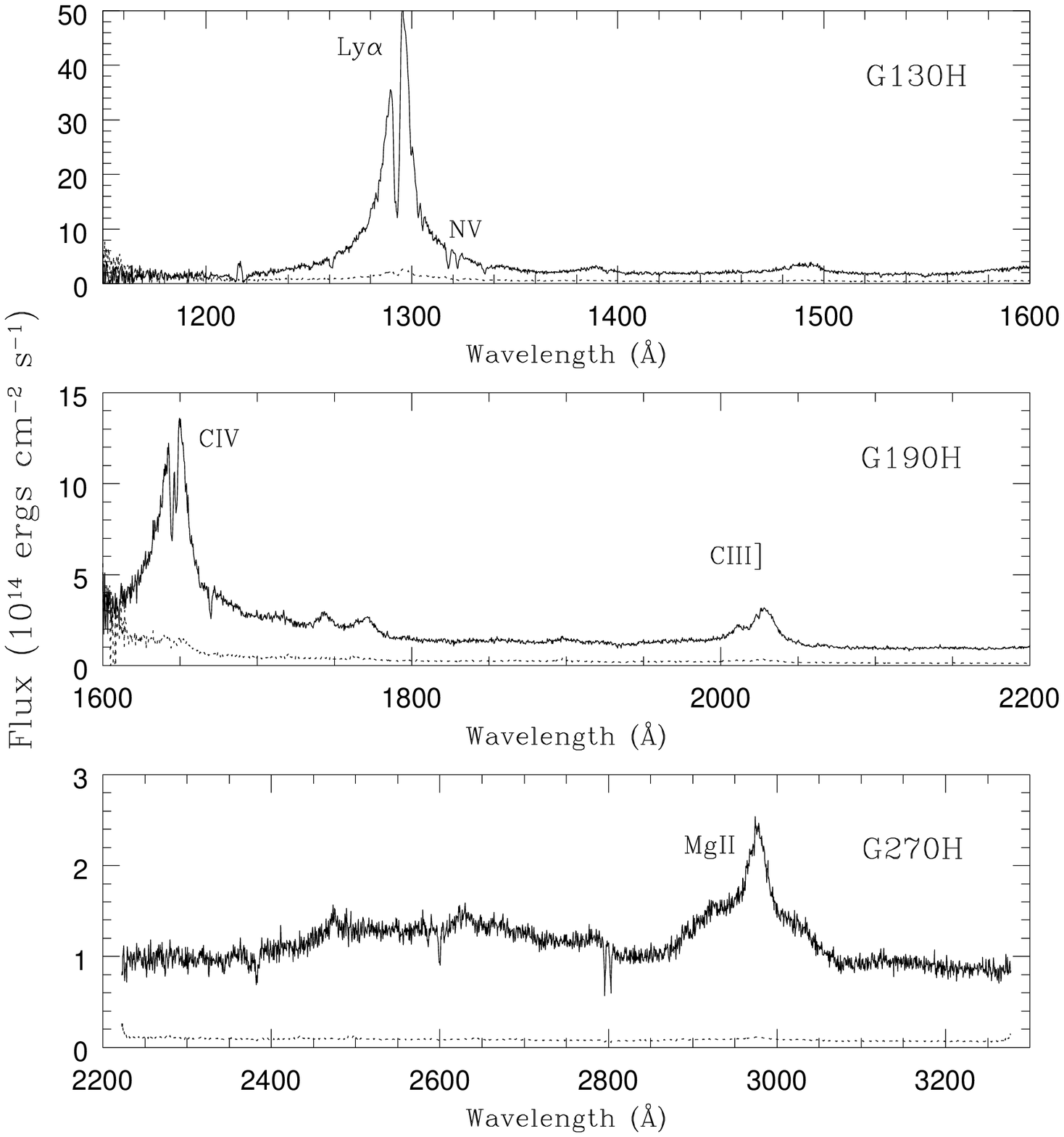}
\end{figure}

\begin{figure}
\epsscale{0.65}
\plotone{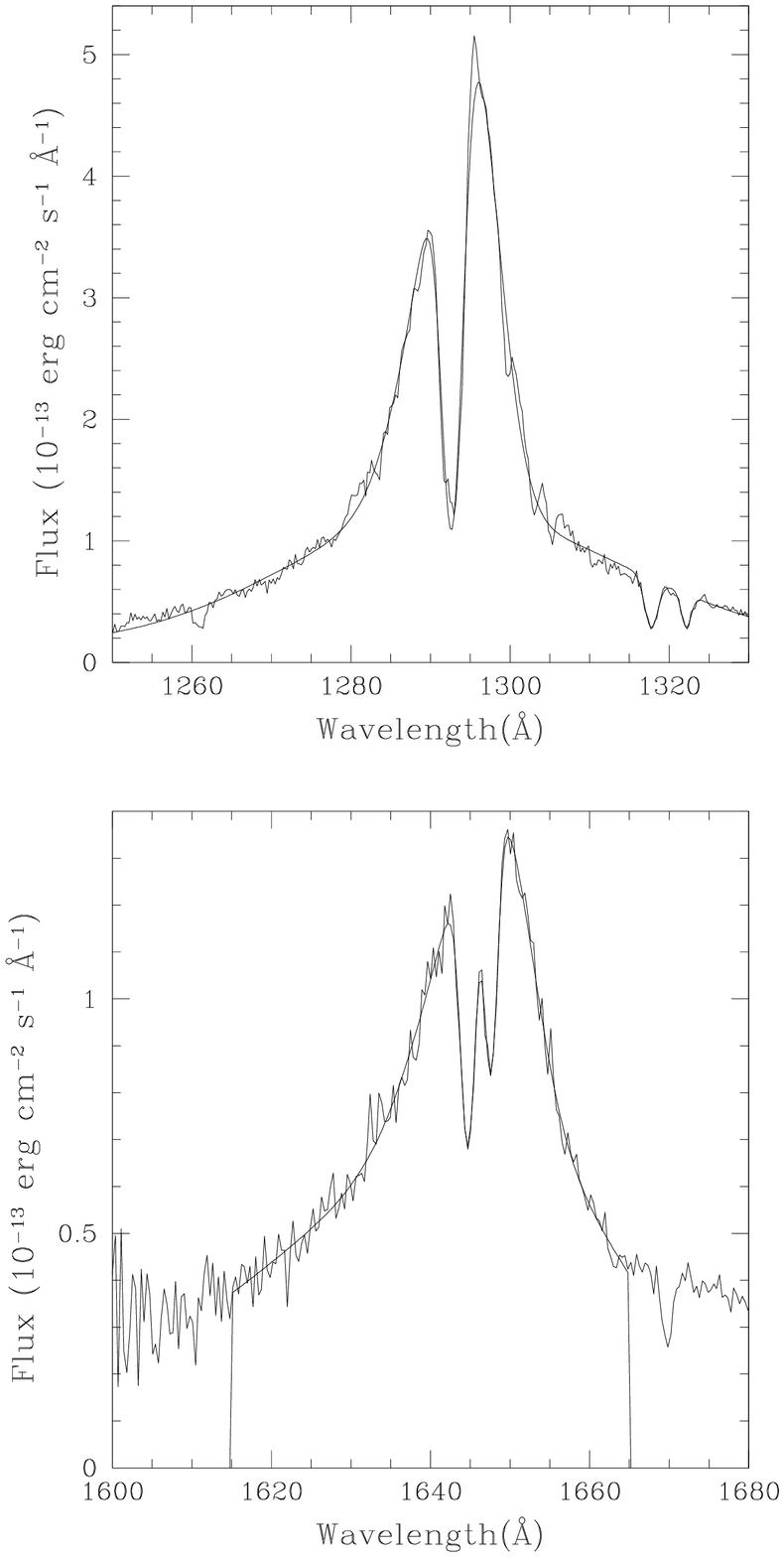}
\end{figure}

\begin{figure}
\epsscale{0.70}
\plotone{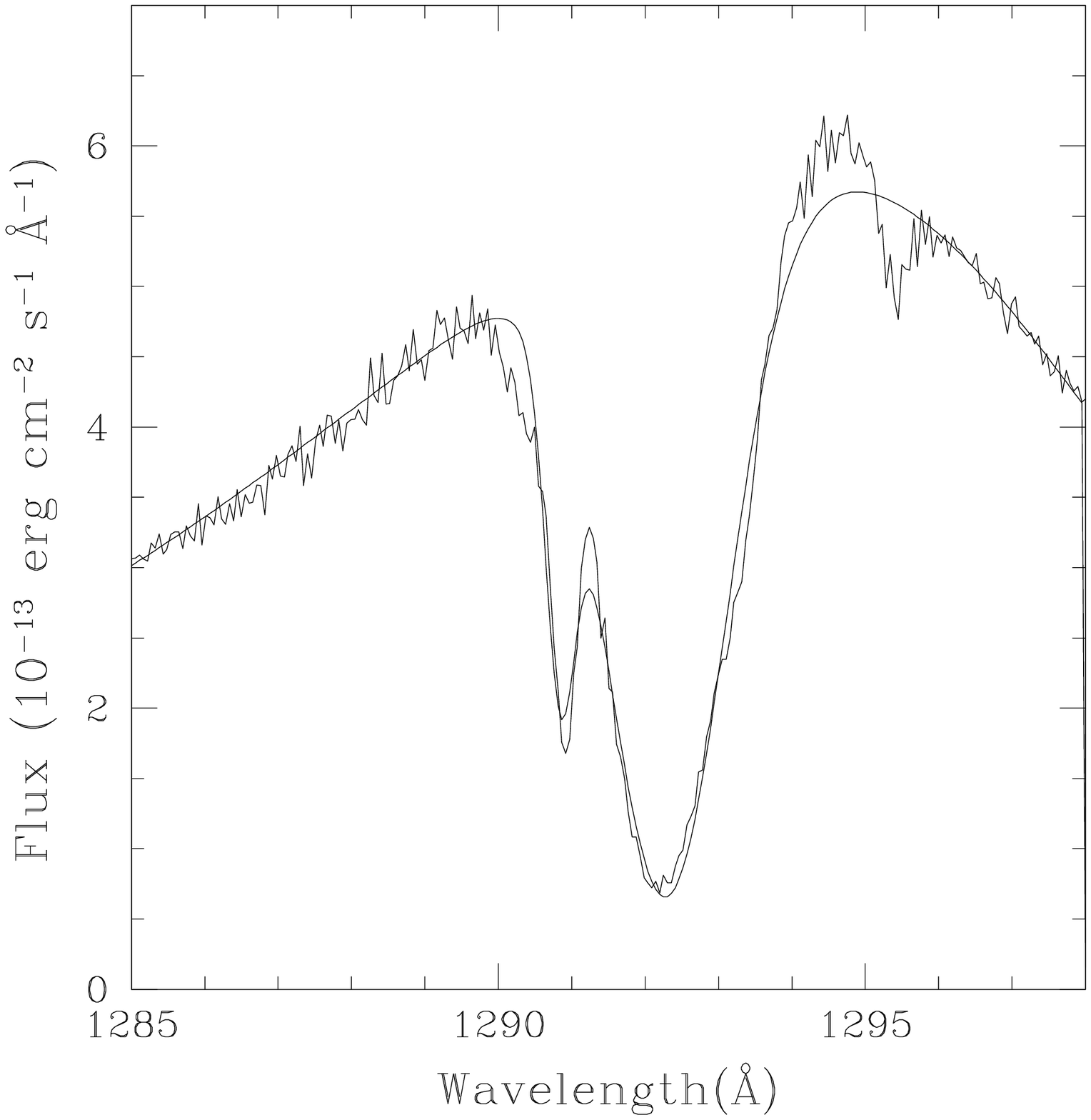}
\end{figure}

\clearpage

%\documentstyle[aasms4,12pt]{article}
%\pagestyle{empty}
%\vspace{-12.0in}

%\begin{document}

%\ptlandscape
\def\lya{Ly$\alpha$}
\def\arcsecpt{\makebox[0pt]{ .}^{\prime\prime}}
\def\arcminpt{\makebox[0pt]{.}^{\prime}}

\begin{deluxetable}{cccccccccc} 
  \tabletypesize{\scriptsize}
\tablewidth{7.0in} 
%\vspace*{-2.0in}
\tablecaption{\bf  Absorption Line Parameters}

\tablehead{
\multicolumn{1}{c}{}  &  \multicolumn{1}{c}{} & \multicolumn{1}{c}{} & 
\multicolumn{1}{c}{}  &  \multicolumn{2}{c}{EW\tablenotemark{a}\ (\AA)} &
 \multicolumn{2}{c}{FWHM (km s$^{-1}$)}  & \multicolumn{2}{c}{$N_{ion}$ (cm$^{-2}$)} \\
 \multicolumn{1}{c}{Line} & \multicolumn{1}{c}{Instrument} &
 \multicolumn{1}{c}{$\lambda_{obs}$ (\AA)}  & \multicolumn{1}{c}{$z$} & 
 \multicolumn{1}{c}{SPLOT}  &  \multicolumn{1}{c}{SPECFIT}  &
 \multicolumn{1}{c}{SPLOT}  & \multicolumn{1}{c}{SPECFIT} & 
 \multicolumn{1}{c}{Observed\tablenotemark{b}} &  \multicolumn{1}{c}{Predicted\tablenotemark{c}} \\
 }
\startdata
Ly$\alpha$1215.7 & FOS-G130H & 1292.6 & 0.0630   &  2.22 &   2.52 $\pm$ 0.08  & 694 & 754 $\pm$ 23  & $>$(3.8 - 4.4) $\times 10^{14}$ & $8.4 \times 10^{14}$   \\
Ly$\alpha$1215.7 & STIS-G140M & 1290.9 & 0.0619   &   --  &   0.26 $\pm$ 0.01  &  -- & 108 $\pm$ 37  & (6.4 $\times 10^{13}$)\tablenotemark{d}  &  \\
Ly$\alpha$1215.7 & STIS-G140M & 1292.3 & 0.0630   &   --  &   1.90 $\pm$ 0.01  &  -- & 475 $\pm$ 24  & (6.4 $\times 10^{14}$)\tablenotemark{d}  &  \\
C IV1548.2 & FOS-G190H & 1644.7 & 0.0623   &  1.07 &   1.09 $\pm$ 0.08   & 434 & 410
$\pm$ 25 & $>$(6 - 7) $\times 10^{14}$ & $1.6 \times 10^{13}$  \\
C IV1550.8 & FOS-G190H & 1647.6 & 0.0624   &  0.64  &   0.80 $\pm$ 0.07 & 399 & 363 $\pm$ 25 &                     &                             \\
N V1238.8 & FOS-G130H & 1317.8 & 0.0637   &  1.02 &   1.14 $\pm$ 0.11  & 382 & 444
$\pm$ 51 & $>$(10 - 11) $\times10^{14}$ & $1.6 \times 10^{14}$         \\
N V1242.8 & FOS-G130H & 1322.2 & 0.0638   &  0.61 &   0.71 $\pm$ 0.10   & 304 & 325
$\pm$ 57 &              &                             \\

\enddata
\tablenotetext{a}{Rest-frame EW.  SPLOT error estimates are 40\% on Ly$\alpha$ and CIV, 10\% on NV.}
\tablenotetext{b}{The FOS values are lower limits from the linear part of the C.O.G.  Two numbers correspond to SPLOT and SPECFIT values.}
\tablenotetext{c}{Prediction from photoionization code CLOUDY using "Table AGN" input continuum.}
\tablenotetext{d}{Curve-of growth measurement.}

\end{deluxetable}
%\end{document}

\end{document}